\font\twelvebf=cmbx12
\font\ninerm=cmr9
\nopagenumbers
\magnification =\magstep 1
\overfullrule=0pt
\baselineskip=18pt
\line{\hfil CCNY-HEP 95/6}
\line{\hfil IASSNS-HEP-95/82}
\line{\hfil RU-95-7-B }
\line{\hfil October 1995}
\vskip .3in
\centerline{\twelvebf A gauge-invariant Hamiltonian analysis for non-Abelian}
\centerline{\twelvebf gauge theories in (2+1) dimensions}
\vskip .4in
\baselineskip=14pt
\centerline{\ninerm DIMITRA KARABALI}
\vskip .05in
\centerline{Physics Department*}
\centerline{Rockefeller University}
\centerline{New York, New York 10021}
\centerline{and}
\centerline{The Institute for Advanced Study}
\centerline{Princeton, New Jersey 08540}
\vskip .4in
\centerline{\ninerm V.P. NAIR}
\vskip .05in
\centerline{Physics Department}
\centerline{City College of the City University of New York}
\centerline{New York, New York 10031.}
\footnote{}{* Present address}
\footnote{}{E-mail addresses: karabali@theory.rockefeller.edu, 
vpn@ajanta.sci.ccny.cuny.edu }
\vskip .5in
\baselineskip=16pt
\centerline{\bf Abstract}
\vskip .1in
The Hamiltonian formulation for a non-Abelian gauge theory in two spatial dimensions is
carried out in terms of a gauge-invariant
matrix parametrization of the fields. The Jacobian for the relevant transformation of
variables is given in terms of the WZW-action for a hermitian matrix field. 
Some gauge-invariant eigenstates of the kinetic
term of the Hamiltonian are given; these have zero charge and exhibit a mass gap.
\vfill\eject
\footline={\hss\tenrm\folio\hss}
\magnification =\magstep 1
\overfullrule=0pt
\baselineskip=18pt
\pageno=2

\def\bp {\bar p}
\def\dag {\dagger}
\def\del {\partial}
\def\bdel{\bar{\partial}}

\def\d {\delta}

\def\zb {\bar{z}}
\def\12 {{\textstyle {1 \over 2}}}
\def\bG {\bar{G}}
\def\H {{\cal H}}
\def\vf {\varphi}
\def\S {{\cal S}}
\def\ra {\rangle}
\def\la {\langle}
\def\Tr {{\rm Tr}}
\def\E {{\cal E}}

\noindent
{\bf 1. Introduction}
\vskip .1in
The importance of non-Abelian gauge theories to our understanding of physical
phenomena is well recognized and needs no reiteration.
While the perturbative aspects of such theories were understood many years ago,
nonperturbative phenomena have been difficult to analyze. Detailed analyses
and calculational techniques are still lacking, eventhough
most of the qualitative features are more or less clear.
Gauge theories in two spatial dimensions, which are simpler than 
their analogues in three spatial dimensions, are 
interesting as a guide to the more realistic case, and not
surprisingly, there have been many 
analyses of such theories starting from the early days [1,2]. 
For a non-Abelian gauge theory 
spontaneously broken to
an Abelian subgroup, one can show the existence of a mass gap or the absence of
massless gauge particles due to the condensation
of monopoles [1]. It is very likely that this feature holds true
for the unbroken theory as well. The subject has recently been receiving renewed 
attention. New gauge-invariant parametrizations, variables similar to gravitational
degrees of freedom as well as extensions of old techniques are being explored [3,4].
In this paper, we introduce a fairly simple matrix parametrization of the gauge fields
which amounts to a particular choice of coordinates on the space of gauge-invariant
configurations and work out the Hamiltonian formulation. This approach is
closest in spirit to the work of Feynman (and also Singer)
who emphasized the importance of the
geometry of the configuration space [5]. The kinetic energy term for the Hamiltonian
is essentially a Laplace operator on the configuration space. Now, 
for a finite-dimensional
compact manifold the Laplacian has a discrete spectrum, the gap between the first and
second
eigenvalues being the analogue of the mass gap. Feynman argued that this picture should
remain qualitatively correct for the infinite-dimensional case of the configuration space
of the (2+1)-dimensional non-Abelian gauge theory. Since the Hamiltonian is a real
operator, the ground state wave function can be chosen real and positive everywhere on 
the configuration space. The excited states, since they have to be orthogonal
to the ground state, must involve nodes. The low-lying excited states can thus
be analyzed by
considering the minimization of the expectation value of the Hamiltonian for wave 
functions with nodes. The potential energy can be minimized by making the magnitude of 
the wave function small. 
The kinetic energy is proportional to the square of the gradient 
(over the configuration space)
of the wave function. One can try to minimize the magnitude of the
gradient by choosing a function for
which the regions where it is positive and where it is negative are, roughly
speaking, as far separated as possible. However, if the gauge-invariant
distance between configurations cannot be made arbitrarily large, the gradient of
the wave function and hence the kinetic energy cannot be made arbitrarily small; 
in other words, one has a mass gap for the theory. This is the essence of Feynman's
arguments, although, of course, distances, expectation values, etc. have to
be understood with regularizations and all that.
By choosing a simple enough
parametrization of the fields, which makes the 
evaluation of the relevant
Jacobian factors and the transformation of the Hamiltonian
quite elementary, we expect to make these intuitive and
qualitative arguments more explicit. 
This is also equivalent to making a special choice of `collective
coordinates' and so our analysis is
similar to some previous work along these lines [6].

For the matrix parametrization we use, the relevant Jacobian factor
is related to the action ${\S}$ for a hermitian Wess-Zumino-Witten (WZW) model [7], i.e.,
a $G^{\bf C}/G$-coset model [8,9], where $G$ is the gauge group and $G^{\bf C}$ is its
complexification. We obtain the Hamiltonian and other generators of the Poincar\'e 
algebra in this parametrization. In view of Feynman's arguments, it is interesting to
consider the spectrum of the kinetic energy term by itself. 
We obtain some
(gauge-invariant) eigenstates of the kinetic energy; 
these correspond to massive states.
The construction of the states involve nonlocal composite operators (in
terms of the gauge potentials) and 
there are arguments which suggest that the norms 
should exist in a suitably regularized sense. We also give arguments regarding 
the completeness of these states.

In the next section, we introduce the matrix parametrization of the gauge fields. Section
3 deals with the reduction of the metric on the space of configurations 
and the Hamiltonian.
In section 4, we discuss the vacuum and other eigenstates of 
the kinetic energy. A short discussion and recapitulation of results is given
in section 5 and
the paper concludes with an appendix where some estimate of
the effect of coordinate singularities is given.
\vskip .1in
\noindent
{\bf 2. The matrix parametrization of fields}
\vskip .1in
We consider a $G= SU(N)$-gauge theory in the $A_{0}=0$ gauge. The gauge potential
can be written as $A_i = -i t^a A_i ^a$, $i=1,2$, where $t^a$ are hermitian 
$N \times N$-matrices which form a basis of the Lie algebra of $SU(N)$ with
$[t^a, t^b ] = i f^{abc} t^c,~~{\Tr} (t^at^b) = {1 \over 2} \delta ^{ab}$. 
The Hamiltonian
can be written as
$$\eqalign{
\H& = T~+V \cr
T&={e^2\over 2} \int d^2 x ~ E_i^a E_i^a \cr
V&= {1\over 2e^2} \int d^2x~ B^a B^a \cr}
\eqno (2.1)
$$
where $e$ is the coupling constant; $e^2$ has the dimension of mass.
Further, $B^a={1 \over 2} \epsilon_{jk}(\partial_j A_k^a - \partial_k A_j^a +f^{abc}
 A_j^b A_k^c )$ and we have the standard commutation rules
$$
\eqalignno {
[A_i^a (x), A_j^b (y)] & = [E_i^a (x), E_j^b (y)] =0 \cr
[A_i^a (x), E_j^b (y)] & = i \delta _{ij} \delta ^{ab} \delta ^{(2)} (x-y) 
&(2.2) \cr} $$
These commutation rules or the corresponding Poisson brackets are given by the
symplectic two-form
$$
\eqalign{
\omega &= \delta \Theta = \int d^2 x ~ \delta E_i^a \delta A_i^a\cr
\Theta &= \int d^2 x ~ E_i^a \delta A_i^a\cr}\eqno(2.3)
$$
We shall use $\delta$ to denote exterior differentiation on the space of gauge
potentials ${\cal{A}} = \{A_i^a (x)\}$. For fields which fall off at spatial
infinity, the spatial manifold can be taken as the Riemann sphere ${\bf C} \cup 
\infty$ with
complex cordinates $z=x_1 - i x_2, ~~ \bar{z} = x_1 +i x_2$. The complex 
components of the potential, viz., $A_{z} = {1 \over 2} (A_1 +i A_2), ~~ 
A_{\bar{z}} =
{1 \over 2} (A_1 -i A_2) = - (A_z)^{\dagger}$, can be parametrized in terms 
of a complex $G^{\bf C}$-, specifically $=SL(N,{\bf C})$-, matrix $M$ as
$$
A_z = -\partial_{z} M M^{-1},~~~~~~ A_{\bar{z}} = M^{\dagger -1} \partial_
{\bar{z}} M^{\dagger}
\eqno(2.4) $$
For a given potential, the matrix $M$ is not uniquely defined; $M$ and $M {\bar V}$,
where ${\bar V}$ is antiholomorphic, lead to the same potential. For Riemann sphere,
the only ( nonsingular or globally defined)
antiholomorphic function being a constant, ${\bar V}$ has to be constant.
This ambiguity can be eliminated by requiring 
$M \rightarrow 1$ as $|x| \rightarrow \infty$. (If we allow singularities in $M$, more 
general choices of ${\bar V}$ are possible. These can be related to the coordinate
singularities of our parametrization of the configuration space;
we shall comment on them in the appendix.) The Green's functions for the $\partial_z$ 
and $\bdel_{\zb}$ can be defined as follows.
$$ \eqalignno{
 \bar{\partial}_x \bar{G} (x,y) ~= \partial _x &G(x,y) ~= \delta ^{(2)} (x-y) \cr
 \bar{G} (x,x') = {1 \over {\pi (z-z')}}, ~~~&~~~~~~
G(x,x') = {1 \over {\pi (\bar{z} - \bar{z}')}} &(2.5) \cr} 
$$
Using these Green's functions, eqs.(2.4) can be inverted to obtain $M$ and $M^\dag$
in terms of $A_z$ and $A_{\zb}$ respectively, at least as a power series
in the gauge potentials.

Using the above parametrization, we
find
$$
\Theta = 2 \int \Tr ~ ( \bar{p} ~ \delta M^{\dagger} M^{\dagger -1} + p M^{-1} \delta
M)
\eqno(2.6) $$
where
$$ \eqalignno{
\bar{p} & = \bar{p}^a t^a =~ 2 ~\partial_{\bar{z}}(M^{\dagger} E_z M^{\dagger -1}
) \cr p & = p^a t^a = - 2 ~\partial_z(M^{-1} E_{\bar{z}} M) = (\bar{p})^{\dagger}
&(2.7) \cr} $$
The symplectic two-form is obtained from (2.6) as
$$
\omega = \delta \Theta = 2 \int \Tr ~ [ \delta \bar{p} ~ \delta M^{\dagger} M^{
\dagger -1} + \bar{p}~ (\delta M^{\dagger} M^{\dagger -1})^2 + \delta p M^{-1}
\delta M - p (M^{-1} \delta M)^2 ]
\eqno(2.8) $$
We want to rewrite the theory in terms of $M$'s. The transformed version of the
commutation rules (2.2) can be obtained from (2.8) as
$$
[F, G] = i ( V_F \rfloor V_G \rfloor \omega) = -i (V_F \rfloor \delta G )
\eqno(2.9) $$
where $V_F$ is the Hamiltonian vector field corresponding to $F$, defined by
$V_F \rfloor \omega = - \delta F $ and the contraction $\rfloor$ is defined by 
$V \rfloor \omega=
V^I \omega _{IJ} \delta \xi ^J $ for $V= V^I {\delta \over {\delta \xi ^I}}, ~
\omega = {1 \over 2} \omega _{IJ} \delta \xi ^I \delta \xi ^J$.

The vector fields $V(\phi),~ \bar{V} (\phi)$ corresponding to $\int p_a \phi ^a$
and $\int \bar{p}_a \phi ^a$ are easily identified as
$$ \eqalignno{
V(\phi) & = v(\phi) -i \int f^{abc} \phi_a p_b {\delta \over {\delta p_c}} 
\cr \bar{V}(\phi) & = \bar{v}(\phi) + i \int f^{abc} \phi_a \bar{p} _b 
{\delta \over {\delta \bar{p}_c}} &(2.10) \cr} $$
where $v(\phi) \rfloor \delta M = M (\phi_a t^a), ~~ \bar{v} (\phi) \rfloor 
\delta M^
{\dagger} = (\phi_a t^a) M^{\dagger}$. Using (2.9) and (2.10), we immediately 
find the commutation rules
$$ 
\eqalign{
[p_a (x), M (y) ] & = -i M (x) t_a ~ \delta ^{(2)} (x-y) \cr
[p_a (x), p_b (y) ] & = f_{abc} p_c(x)~ \delta ^{(2)} (x-y) \cr 
[\bar{p}_a (x), M^{\dagger} (y) ] & = -i t_a M^{\dagger} (x)~ \delta ^{(2)} 
(x-y) \cr
[\bar{p}_a (x), \bar{p}_b (y) ] & = -f_{abc} \bar{p}_c (x) ~ \delta ^{(2)} (x-y) 
\cr
[M(x), M (y) ] & = ~ [M^{\dagger} (x), M^{\dagger} (y)]~=~ [M(x), M^{\dagger} 
(y)] = 0 \cr
[p_a (x), M^{\dagger}] & =~ [\bar{p}_a (x), M(y)]~=~ [p_a (x), \bar{p}_b (y)] 
 =0 \cr} \eqno(2.11)
$$

The $SL(N,{\bf C})$-matrix $M$ can be parametrized in terms of a complex field 
$\theta _a (x)$. We write
$$ \eqalignno{
M^{-1} \delta M & = \delta \theta ^a R_{ab} t^b \cr
\delta M^{\dagger} M^{\dagger -1} & = \delta \bar{\theta} ^a R^{*} _{ab} t^b 
&(2.12) \cr} 
$$
The definition of the potentials leads to
$$\eqalign{
\delta A_{z~k} &= -i {\cal E}_{ka} \partial_z (\delta \theta^b R_{ba})\cr
\delta A_{{\bar z}~k} &= ~i {\cal E}_{ak}^\dagger \partial_{\bar z}(\delta {\bar \theta}^b
R^*_{ba})\cr}\eqno(2.13)
$$
where
$$
{\E}_{ka} = 2 ~\Tr (t_k M t_a M^{-1}),~~~ {\E}^{\dagger} _{ak} = 2 ~\Tr (t_a M^{\dagger}
t_k M^{\dagger -1}) \eqno(2.14)
$$

The commutation rules (2.11) can be realized by the following differential
operator representation for $p_a,~\bar{p}_a$.
$$ \eqalignno{
p_a (x) & = -i R_{ab}^{-1} (x) ~ {\delta \over {\delta \theta ^b (x)}} \cr
\bar{p}_a (x) & = -i R_{ab}^{*-1} (x)~ {\delta \over {\delta \bar{\theta} ^b 
(x)}} &(2.15) \cr}  
$$
The components of the electric field can be written as
$$ \eqalignno{
E_{z ~k} & = ~{i\over 2} {\E}^\dagger_{ak}\int _y \bar{G} (x,y) \bar{p}_a (y) \cr
E_{\bar{z}~k}  & = -{i\over 2} {\E}_{ka}\int _y G(x,y) p_a(y) &(2.16) \cr} 
$$
These expressions have the correct ordering of the operators; this can be 
seen by deriving them from a change of variables. Thus
$$\eqalign{
E_{z ~k}(x)\Psi(M, M^\dagger )&= -{i\over 2} {\delta \Psi \over 
\delta A_{\zb ~k}(x)}
= -{i\over 2} \int_y {\delta \Psi \over \delta {\bar \theta}^a (y)} {\delta {\bar \theta}^a (y)\over
\delta A_{\zb~k}(x) }\cr
&= -{1\over 2}\int_y {\delta \Psi \over \delta {\bar \theta}^a}R^{*-1}_{ba}(y) {\bar G}(y,x)
{\cal E}_{bk}^\dagger (x)\cr}\eqno(2.17)
$$
where we have used (2.13).

The new variables $M$ and $M^\dag$ are not gauge-invariant.
Gauge transformations act on $M, M^{\dagger}$ as $M \rightarrow g M \simeq (1 +
i \phi ^a t_a) M, ~ M^
{\dagger} \rightarrow M^{\dagger} g^{-1} \simeq M^{\dagger} (1- i \phi ^a t_a), 
~~g(x) \in SU(N)$. The theory can be written in terms of the gauge-invariant
combination $H= M^\dag M$, which can be considered as the fundamental dynamical
variable. 
From the commutation rules (2.11), we obtain
$$\eqalign{
[p_a (x), H (y) ] & = -i H (x) t_a ~ \delta ^{(2)} (x-y) \cr
[\bar{p}_a (x), H(y) ] & = -i t_a H(x)~ \delta ^{(2)} (x-y) \cr}\eqno(2.18)
$$
When acting on functions of $H$, $p_a$ and ${\bar p}_a$ can be represented as
$$\eqalign{
p_a(x) &= -i r^{-1}_{ab}(x) {\delta \over {\delta \vf^b (x)}}\cr
{\bar p}_a(x) &= -i r^{*-1}_{ab}(x) {\delta \over {\delta \vf^b (x)}}~=
K_{ab}(x)p_b(x) \cr}
\eqno(2.19)
$$
where we have defined $H$ in terms of the real parameters $\vf^a$ and 
$$
H^{-1} \delta H~= \delta \vf^a r_{ab}~ t^b \eqno(2.20)
$$
The Gauss law operator
or the generator of gauge transformations can be obtained as
$$
G(\phi) = i \int \phi _k ~ ({\E}_{ka} p_a - {\E}^{\dagger} _{ak} \bar{p}_a) \eqno(2.21)
$$
From the
commutation rules, we  see that $p_a, \bar{p} _a $ are gauge-invariant operators, i.e.,
$$
[G_k, p_a] = [G_k, \bar{p}_a] = 0 \eqno(2.22)
$$
We have already used this fact in representing $p_a,~ {\bar p}_a$
as derivatives with respect to $\vf^a$ in (2.19).
The differential operator representations
(2.15) which are appropriate when acting on
functions of $M$ and $M^\dag$ 
naturally go over to the 
representation (2.19) for action on functions of $H$.

The magnetic field can be written as
$$ 
B= B^a t_a = -2 ~M[\partial ( H^{-1}{\bar \partial} H) ] M^{-1}
\eqno (2.23) 
$$
where $\partial \equiv \partial_z,~
{\bar \partial}\equiv \partial_{\bar z}$.  
The potential term in the Hamiltonian
can thus be written as
$$
V~={2\over e^2}\int d^2x~[ \partial (H^{-1} \bar{\partial} H )_a 
\partial (H^{-1} \bar{\partial}
H)_a] \eqno(2.24) 
$$
Using the expressions (2.16) for the electric fields, the kinetic term $T$ can be written as
$$
T~= {e^2\over 2} \int d^2 x ~ \int _{y,z} \bar{G} (x,y) \bar{p}_a (y) 
K_{ab}(x)  G(x,z)
p_b (z) \eqno(2.25)
$$
where
$$ 
K_{ab} = 2 ~\Tr (t_a H t_b H^{-1}) ~= {\E}^\dagger_{ak}{\E}_{kb}\eqno(2.26) 
$$
$K_{ab}$ is the adjoint representation of the matrix $H$; we also have
$K^T K =1~, {\E}^T {\E}=1, ~{\E}^{\dagger T}{\E}^\dagger =1$, 
where the superscript $T$ denotes the
transpose.

The kinetic term of the Hamiltonian as given in (2.25) is still essentially 
classical since we
have not included the corrections from 
the Jacobians which arise in writing the inner product of wavefunctions in
terms of $M$, and eventually $H$. 
The kinetic energy $T$, as given by (2.1), involves the Laplacian on ${\cal{A}}$, 
since $E_i^a (x) = -i {\delta \over {\delta A_i^a(x)}}$. 
Upon changing to $M, M^{\dagger}$, this
should be the Laplacian expressed in terms of $M, M^{\dagger}$ (or $H$).
Expression (2.25) gives only one of the terms in the Laplacian; the remaining terms,
which involve the determinant of the metric, will arise from the Jacobian in the 
inner product for wave functions.

The physical states are gauge-invariant and obey $G |{\rm phys}\ra =0$. They can be 
represented as functions of $H$.
Further, 
from the discussion following eq.(2.4), $H$ and $V H{\bar V}$, where $V$ is a constant
$SL(N,{\bf C})$-matrix, describe the same
physical configuration; thus we must require that physical states, and correspondingly
observables, should be invariant under $H\rightarrow VH{\bar V}$.
\vskip .1in
\noindent
{\bf 3. The reduction of the metric, the Hamiltonian and Poincar\'e invariance}
\vskip .1in
The metric on the space of gauge potentials ${\cal{A}}$ is given by
$$
ds^2 = \int \delta A_i^a \delta A_i^a = -8 \int \Tr (\delta A_z \delta A_
{\bar{z}}) \eqno(3.1) 
$$
The wavefunctions 
have the inner product
$$
\la 1|2\ra = \int d\mu ({\cal{C}})~ \Psi_1^* \Psi_2 \eqno(3.2)
$$
where $d\mu ({\cal{A}})$ is defined by the metric (3.1) and $d\mu({\cal{C}})$, the
measure on the space of physical configurations ${\cal{C}}$, is obtained from
$d\mu ({\cal{A}})$ by factoring out the volume of gauge transformations. 
Formally
$$
d\mu ({\cal{A}}) = \prod _x dA_z (x) dA_{\bar{z}} (x) \eqno(3.3)
$$
More properly (3.3) should be understood with appropriate regularization.

We first express the measure $d\mu ({\cal C})$ in the inner product (3.2)
in terms of $H$. Using the definition of $A_z,~A_{\bar z}$, we have 
$$ 
ds^2 = 8 \int \Tr [ D_z (\delta M M^{-1}) D_{\bar{z}} ( M^{\dagger -1} \delta
M^{\dagger})] \eqno(3.4)
$$
where $D_i \phi = \partial _i \phi + [A_i, \phi]$. The metric 
for $SL(N,{\bf C})$-matrices can be written as
$$ \eqalignno {
ds^2 & = 8 \int \Tr [(\delta M M^{-1}) (M^{\dagger -1} \delta M^{\dagger})] \cr
& = 2 \int \delta \bar{\theta} (R^* K R^T) \delta \theta + \delta \theta (R K^T
R^{*T}) \delta \bar{\theta} &(3.5) \cr}
$$
The Haar measure $d\mu (M, M^{\dagger} )$ is the volume associated 
with this 
metric. Explicitly, $d\mu (M, M^{\dagger}) = (\det R R^*) [ \delta 
\theta \delta
\bar{\theta}]$.
From eqs.(3.4, 3.5) we see immediately that [8,9,10]
$$
d\mu ({\cal{A}}) = [dA_z dA_{\bar{z}}] = (\det D_z D_{\bar{z}}) d\mu 
(M, M^{\dagger} ) \eqno(3.6) 
$$
$D_z$ and $D_{\bar{z}}$ are chiral covariant derivatives in two dimensions and 
their determinants are well known. The gauge-invariant evaluation of these 
determinants gives [8-11]
$$\eqalign{
\det (D_z D_{\bar{z}}) &= \exp [2 c_A {\S} (H)] ~\sigma \cr
\sigma &=\left[ {\det ' (\partial {\bar \partial})
\over \int d^2x }\right]^{{\rm dim}G} \cr}
\eqno(3.7)
$$
where $c_A$ is the quadratic Casimir of the adjoint representation, viz., $c_A 
\delta _{ab} = f_{amn} f_{bmn}$; ${\rm dim}G = (N^2-1)$ is the dimension of the gauge 
group
$SU(N)$. The prime on the determinant of 
$(\partial {\bar \partial})$ and the division by the volume of space denote the 
factoring out of the constant or zero mode of $(\partial {\bar \partial})$.
${\S} (H)$ is the Wess-Zumino-Witten (WZW) action for $H$.
$$
{\S} (H) = {1 \over {2 \pi}} \int \Tr (\partial H \bar{\partial} H^{-1})
+{i \over {12 \pi}} \int \epsilon ^{\mu \nu \alpha} \Tr ( H^{-1} \partial _{\mu}
H H^{-1} \partial _{\nu}H H^{-1} \partial _{\alpha}H) \eqno(3.8) 
$$
As is usual for the WZW action, we write the second term in $\S$ in terms of an
extension of $H$ into a three-dimensional space with space as the boundary. Actually,
for hermitian matrices such as $H$, this is unnecessary and the second term can be 
integrated and written as an integral over space only [9,12].

Consider now the reduction of $d\mu (M, M^\dagger )$ by factoring out the volume of 
gauge transformations. Since gauge transformations act as $M\rightarrow g M$, $g\in G$,
the physical degrees of freedom belong to $G^{\bf C}/G$ and we need to construct 
the volume element on 
$G^{\bf C}/G$. The metric obtained from (3.5) by factoring out gauge transformations
is
$$
ds^2 ~= 2 \int ~\Tr(H^{-1}~\delta H )^2 \eqno(3.9)
$$
The volume corresponding to this is given by
$$
d\mu (H) ~= \det r  ~[\delta \vf ] \eqno(3.10)
$$

A simple way to see this result is as follows.
$\rho \equiv ((M^\dagger)^{-1} d M^\dagger ~+ dM M^{-1})$ is a 
differential form on $G^{\bf C}$ 
which transforms as
$\rho \rightarrow g \rho g^{-1}$ under $M\rightarrow g M$. 
Thus $\Tr ( \rho ^n)~= \Tr (H^{-1}dH )^n $ are differential forms on
$G^{\bf C}/G$. The volume element is given by the differential form of maximal
degree, i.e., for $n= (N^2 -1)$. This is easily seen to be $\det r [d\vf ]$. 
For matrices which are functions of the spatial coordinates, as in our case, we have
the product over the spatial points as well [9,10]; this is understood in eq.(3.10).

Using eqs.(3.6-3.10), we can write the inner product as
$$
\la 1|2\ra = \int d\mu ({\cal C})~ \Psi^*_1 \Psi_2
= \int  \sigma ~d\mu (H) e^{2c_A {\S} (H) }~\Psi^*_1 \Psi_2 \eqno(3.11)
$$

With the inner product (3.11) and the representation (2.19),  
the expression for the kinetic energy $T$ can now be obtained as follows.
We can write
$$ \eqalignno{
\la 1|T|2\ra &=
 e^2 \int d\mu ({\cal C})~ [(E_{\bar{z}}^a \Psi_1)^* (E_{\bar{z}}^a \Psi_2) + 
(E_z^a \Psi_1)^* (E_z^a \Psi_2)] &(3.12) \cr
&= {e^2\over 4} \int \sigma
 d\mu (H) e^{2 c_A {\S} (H)}\left[(Gp_a \Psi_1)^* K_{ab} 
(G p_b \Psi_2) + (\bar{G} \bar{p} _a \Psi_1)^* K^{T}_{ab} 
(\bar{G} \bar{p}_b \Psi_2)\right]  
\cr}
$$
where we have used the change of variables procedure as in (2.17) and $(Gp_a) (x) = \int _y G(x,y) p_a (y) $.
We do a partial integration using the easily verified property 
$$
{\delta \over
{\delta \vf_b}} [(r^{* -1} _{ab} \det r^*)f] = r^{* -1} _{ab} (\det r^*)
{\delta \over{\delta \vf_b}}f.\eqno(3.13)
$$
This gives
$$
\la 1|T|2\ra =
{e^2\over 4}\int  \sigma d\mu (H)~ \Psi_1^* [
(\bar{G}\bar{p}_a) ( K_{ab}e^{2 c_A \S} G p_b ) + (G p _a) K^{T}_
{ab} e^{2 c_A \S}(\bar{G} \bar{p}_b) ]  \Psi_2 \eqno(3.14)
$$
We can thus identify the operator $T$ as
$$
T = {e^2\over 4} e^{-2 c_A \S} ~ \bigl[ ~(\bar{G}\bar{p}_a) K_{ab} e^{2 c_A \S} 
(G p_b) + (G p_a)  K^{T}_
{ab} e^{2 c_A \S}(\bar{G} \bar{p}_b) \bigr] \eqno(3.15)
$$
$T$ is of course self-adjoint by construction.
Notice that if we write $T=- {e^2\over 2} \Delta $, 
then $\Delta$ has the form of a 
Laplacian.
The above method is standard for working out the form of the kinetic energy under change
of variables. It has been used for collective coordinates and also in the context 
of $A_0=0$ gauge quantization of
a gauge theory (with  
parametrizations of the fields different from ours) [6]. The form of the metric 
is really not explicitly used in this derivation; given the Jacobian and the change of 
variables formula (2.17), eq.(3.15) follows. If required, 
the inverse of the metric can actually be read off
from this equation by comparison with the standrad form of the Laplacian.

Given the inner product (3.11), we see that it is convenient to define the
wavefunctions
$$ \Phi = e^{c_A \S} \Psi \eqno (3.16) 
$$
The inner product is now
$$
\la 1|2\ra  = \int \sigma d\mu (H) ~\Phi ^* _1 \Phi _2 \eqno (3.17)
$$
and the Hamiltonian can be written as
$$ 
{\H} = {e^2 \over 4} \int ( P_a^{\dagger} K_{ab} P_b + Q_a^{\dagger} K^{T}_{ab}
Q_b) + V \eqno (3.18) 
$$
where
$$ 
\eqalign{
P_a & = \int G (p_a - c_A~ p_a \S) \cr
P_a^{\dagger} & = \int \bar{G} (\bar{p}_a + c_A ~\bar{p}_a \S) \cr
Q_a & = \int \bar{G} (\bar{p}_a - c_A~ \bar{p}_a \S) \cr
Q_a^{\dagger} & = \int G (p_a + c_A~ p_a \S)  \cr}\eqno(3.19)
$$
Explicitly
$$ p_a {\S} = -{i \over {\pi}}  \Tr [ t_a \partial (H^{-1} \bar{\partial} H)]~,
~~~~~~\bar{p}_a {\S} = -{i \over {\pi}}\Tr [ t_a \bar{\partial} (\partial H H^{-1})] 
 \eqno (3.20) 
$$

One can expand ${\H}$ in terms of $p_a, {\bar p}_a$ and reorder terms 
bringing the derivatives to the right end. 
In doing so we encounter the commutator $[{\bar G}{\bar p}_a(x), K_{ab}(x)]$ which is
singular, since the operators are at the same point. This can be evaluated as follows.
Keeping in mind the relation ${\bar p}_a =K_{ab}p_b$ which is required by Gauss law,
$$\eqalign{
\int_{z}[\bG (x,z) \bp_a (z), K_{ab} (x)] &= \int_{z} \bG (x,z)[ K_{ac}(z) p_c(z), K_{ab} (x) ]\cr
&= \int_{z} \bG(x,z) f_{cbm}K_{ac}(z) K_{am}(x) \Delta(x,z)\cr
&=\int_{z} f_{cbm}\bG(x,z)\left\{ \delta_{cm} -(K^T\partial K)_{mc} (x-z)+...
\right\}\Delta(x,z)\cr
&= i{ c_A\over \pi}(H^{-1}\partial H)_b (x) \cr}
\eqno(3.21)
$$
where $\Delta (x,z)$ is a regularization of $\delta^{(2)} (z-x)$; it can be considered as a
narrow Gaussian function of $(z-x)$. Since it has support only for $z\approx x$, we have
expanded $K_{ac}(z)$ around $x$. The result (3.21) then follows
in the limit $\Delta (z,x) \rightarrow \delta^{(2)} (z-x)$. This evaluation of the commutator
can also be checked with a Pauli-Villars regulator for 
the Green's functions in (3.18) or (3.21)
as well as against the operator
equations of motion. Using (3.20), we may write this result as
$$
\int \bG (x,z)\bp_a(z) K_{ab}(x) = \int K_{ab}(x) \bG(x,z) \bp_a(z) ~ 
-2c_A K_{ab}(x) (\bG \bp_a ){\cal S}\eqno(3.22)
$$
With this result and a similar one for $(Gp)_a K^T_{ab}$ we get 
for the kinetic energy operator
$$
T~=~ {e^2\over 2} \int K_{ab} Q_a P_b \eqno(3.23)
$$

We have made transformations and reductions of variables in the Hamiltonian framework
and since some of these involve singular products, a useful conistency check is the
Poincar\'e invariance of the theory. We have explicitly checked that the
generators
$$ \eqalignno{
{\cal P}^{\mu}  & = \int d^2 x ~T^{0 \mu} \cr
{\cal J}^{\mu \nu}  & = \int d^2 x ~( x^{\mu} T^{0 \nu} - x^{\nu} T^{0 \mu})  &(3.24) \cr}
$$
do satisfy the Poincar\'e algebra, i.e.,
$$ \eqalignno{
[ {\cal P}^{\mu},~{\cal P}^{\nu}]  & =0  \cr
i [{\cal P}^{\mu},~{\cal J}^{\nu \lambda}] & = g^{\mu \lambda}{\cal P}^{\nu} - 
g^{\mu \nu}{\cal P}^{\lambda} \cr
i [{\cal J}^{\mu \nu},~{\cal J}^{\kappa \lambda} ] & = g^{\nu \lambda} 
{\cal J}^{\mu \kappa } - 
g^{\mu \lambda} {\cal J}^{\nu \kappa} + g^{\mu \kappa} {\cal J}^{\nu \lambda}
 - g^{\nu \kappa} {\cal J}^{\mu \lambda} &(3.25) \cr}
$$ 
In our case, the densities for the energy and momentum are given by
$$\eqalign{
T^{00} (x) & = {e^2 \over 4} [ P^{\dagger} _a  (x) K_{ab} (x) P_b (x) +
 Q^{\dagger} _a (x) K_{ba} (x) Q_b (x)] + {2 \over {e^2}} \partial (H^{-1} 
\bar{\partial}H)_a  \partial (H^{-1} \bar{\partial}H)_a \cr
T^{0z}(x) &= -{1\over 4} \bigl[ \bdel (\del H H^{-1})_a (P_a^\dag +Q_a ) +
 (P_a^\dag +Q_a )\bdel (\del H H^{-1})_a\bigr] \cr
T^{0\zb}(x) &= -{1\over 4}\bigl[ \del (H^{-1}\bdel H )_a (P_a +Q_a^\dag ) +
(P_a +Q_a^\dag )\del (H^{-1}\bdel H)_a \bigr]\cr } \eqno(3.26)
$$
It is straightforward to verify that ${\cal P}^i$ and ${\cal J}^{12}$ are the generators 
of translations and rotation respectively, namely,
$$ \eqalignno{
[\Phi (x),~{\cal P}_{i}] & = i \partial _i  \Phi (x) \cr
[\Phi (x),~{\cal J}^{12}] & = 
[S_{\Phi} + i (x_1 \partial _2 - x_2 \partial _1) ] \Phi (x)  
& (3.27) \cr }
$$
where $\Phi (x)$ denotes any function of the fields of the theory.
$S_{\Phi}$ is the spin carried by $\Phi (x)$; 
for example $S_{P^{\dagger}} = S_{Q} =1,~ S_{P}=S_{Q^{\dagger}}=-1,~ S_{K_{ab}} =0$. 

Given eq.(3.27), along with the fact that $T^{00}$ has no explicit coordinate dependence, 
the other necessary and sufficient condition for Poincar\'e invariance is the 
Dirac-Schwinger condition
$$
[T^{00} (x),~ T^{00} (y)] = i (T^{0i} (x) +T^{0i} (y)) \partial_{x}^{i} 
\delta (x-y) \eqno(3.28) 
$$
We have explicitly checked eqs.(3.27,3.28) for the operators (3.24,3.26) of 
the reduced theory.
The commutation rules for the operators $P_a,Q_a, P_a^\dagger ,Q_a^\dagger$, useful for
checking these equations, 
can be worked out from (2.11) and are the following.
$$
\eqalign{
[P_a(x), P_b(y)] &= -f_{abc} G(x,y) \bigl( P_c(x)- P_c(y) \bigr) \cr
[P_a^\dag (x), P_b^\dag (y) ] &= f_{abc} {\bar G}(x,y) \bigl(P_c^\dag (x)- P_c^\dag (y)
\bigr) \cr
[P_a(x), P_b^\dag (y)]& = {c_A\over \pi} K_{ba}(x) \delta^{(2)} (x-y)\cr
[Q_a(x),Q_b(y)] &= f_{abc} {\bar G}(x,y) \bigl( Q_c(x)-Q_c(y) \bigr) \cr
[Q_a^\dag (x), Q_b^\dag (y)]&= -f_{abc}G(x,y)\bigl( Q_c^\dag (x)- Q_c^\dag (y) \bigr)\cr
[Q_a(x), Q_b^\dag (y)] &= {c_A\over \pi} K_{ab}(x) \delta^{(2)} (x-y)\cr
[P_a(x), Q_b(y)]&=[P_a^\dag (x), Q_b^\dag (y)] =0\cr
[P_a(x), Q_b^\dag (y) ] &= -{c_A\over \pi} \delta_{ab} {\bar \partial}_y 
G(x,y)~+f_{abc}G(x,y)Q_c^\dag (y)\cr
&~~~~~~-f_{abc} G(x,y) P_c(x)\cr
[P_a^\dag (x), Q_b(y) ] &= {c_A\over \pi} \delta_{ab} \partial_y {\bar G}(x,y) ~-f_{abc}
{\bar G}(x,y) Q_c(y)\cr
&~~~~~~+f_{abc}{\bar G}(x,y) P_c^\dag (x) \cr }\eqno(3.29)
$$
\vskip .1in
\noindent {\bf 4. Eigenstates of the kinetic energy}
\vskip .1in
The total volume of the space ${\cal{C}}$ of gauge-invariant configurations is 
given by ${\cal V}=\sigma Z$, where
$$ Z = \int  d\mu (H) e^{2c_A \S} \eqno (4.1) 
$$
$Z$ is the partition function on the Riemann sphere of a 
$G^{\bf C}/G$-WZW
model. This partition function is finite, modulo regularization of the Laplacian
$\partial \bar{\partial}$, and can be written as [10]
$$ Z = \big[ {{({\rm det}' \partial \bar{\partial})} \over {\int d^2x}} \big] ^
{-{\rm dim}G} \eqno (4.2) 
$$
which gives ${\cal V}=1$.
For an Abelian theory with $c_A=0$, $Z$ is infinite. 
(If we make a mode decomposition of $H$ or $\vf^a$ over the eigenmodes
of the Laplacian $\partial \bdel$, the integration over the amplitude of 
each mode is finite for the non-Abelian case because of
the exponential; the divergence
arises from the infinity of modes and can be regularized by truncation
to a finite number of modes. For the Abelian case, the integration for each
mode is divergent.) The finiteness of $Z$, albeit
modulo regularization, suggests that the Laplace operator $\Delta$ on 
${\cal{C}}$ should have a mass gap. Thus the diagonalization of the kinetic
energy operator $T$ by 
itself could give us some insight into the origin of the mass gap.
The potential term $V$ involves more derivatives and inverse powers of $e^2$
and for states for which the momenta are small compared to $e^2c_A$, we may
expect that it
can be treated as a perturbation. Of course, neglecting gradient terms and
making such an `ultralocal' approximation [3] will be consistent only if we already have
a mass term or, in our case, if the kinetic
term by itself leads to a mass gap of order $e^2c_A$.

For the purpose of looking for eigenstates, it is actually convenient to go back to the
wavefunctions $\Psi$ rather than the redefined wavefunctions $\Phi$. As an operator on
$\Psi$'s, the kinetic energy can be written, taking account of (3.22), as
$$\eqalignno{
T~&=~{e^2\over 2} \int ~e^{-2 c_A \S}(\bar{G}\bar{p}_a) K_{ab} e^{2 c_A \S} 
(G p_b) &(4.3)\cr
&=~ {e^2\over 2} \int K_{ab} (\bG \bp_a) (G p_b) &(4.4)\cr}
$$
The inner product is given by (3.11). 
$T$ is a manifestly positive operator and its
ground state or vacuum state is defined by
$T ~|0\ra = 0$. 
The solution of this condition is $\Psi_0 =$ constant, say ${\cal N}$.
This may seem somewhat trivial;
the key point, however, is that this function is normalizable with appropriate
regularizations.
The normalization integral is given by ${\cal V}$ in (4.1).
We can choose ${\cal{N}}$ to obtain $\la 0|0\ra = 1$; with the regularization
as in (4.1,4.2), ${\cal N}=1$.

The matrix elements of the operators and the normalization integrals for 
wavefunctions are given by expectation values or correlation functions
of the hermitian $G^{\bf C}/G$-WZW model. 
These can be obtained by analytic continuation of
the correlation functions of the corresponding unitary WZW model [9]. The 
$G^{\bf C}/G$-analogues of the 
level number $k$  and the renormalized level $\kappa =k+c_A$ of the unitary WZW model are
$-(k+2c_A)$ and $-(k+c_A)=-\kappa$. The correlators can be obtained by the continuation
$\kappa \rightarrow -\kappa$. In our case, since we have only $e^{2c_A\S}$, rather than
$e^{(k+2c_A )\S}$, we must also
take $k\rightarrow 0$. 
In the unitary case, correlators involving primary fields belonging to the 
nonintegrable representations of the current algebra vanish. The corresponding statement
for the hermitian model is that such correlators become infinite or undefined [9,10].
In our case, since $k\rightarrow 0$, only the identity and its current algebra
descendents have 
well-defined correlators. (The divergence of correlators of other operators
has to do with $k\rightarrow 0$, not coincidence of arguments.)
We must thus conclude that the wavefunctions for the higher states can be taken as
functions of the current 
$$
J_a ~=~ {c_A \over \pi} (\partial H ~H^{-1})_a \eqno(4.5)
$$

The arguments given above are the conformal field-theoretic reason for the currents
being the quantities of interest. The result, however, is not surprising since the
Wilson loop operator can be written in terms of $H$ as
$$
W(C)~=~ \Tr~ P \exp\left( - \oint_C dz~\partial H H^{-1}\right) \eqno(4.6)
$$
The Wilson loop operators form a complete set and hence the currents $J_a$ should suffice 
to generate the gauge-invariant states.

Consider now a wavefunction
$$
\Psi_a =\int J_a(x) f(x) \eqno(4.7)
$$
for some c-number function $f(x)$. From (4.3) we get
$$
\eqalign{
T \Psi_a ~&= ~{e^2\over 2} \int e^{-2 c_A \S}(\bar{G}\bar{p}_m) K_{mn} e^{2 c_A \S} 
(G p_n) \Psi_a \cr
&=~ i{e^2c_A\over 2 \pi} \int_{z,x} e^{-2 c_A \S} (\bar{G}\bar{p}_m)(z) K_{mn} (z)e^{2 c_A \S} K_{an}(x) 
\delta (z-x)f(x)\cr
&=~ i{e^2c_A\over 2 \pi} \int e^{-2 c_A \S}(\bG \bp _a)(x)  e^{2 c_A \S} f(x)\cr
&=~ \left(e^2 c_A\over 2\pi\right)\Psi_a =~m\Psi_a\cr}\eqno(4.8)
$$
where $m= e^2c_A/2\pi $. This shows that $J_a$ effectively behaves as an eigenfunction
of eigenvalue $m$. The result (4.8) can also be obtained from the version (4.4) of $T$
using the relation
$$
[ K_{mn}(z) (\bG \bp_m)(z) ,K_{an}(x) ]\vert_{z \rightarrow x}~= -i J_a (x) \eqno(4.9)
$$

Of course, $J_a$ by itself is not an acceptable eigenfunction
since it does not have invariance under $H \rightarrow V H {\bar V}$. We can 
construct eigenfunctions with this invariance by taking products of $J$'s. The state
$$
\Psi_2(J) ~=~ \int [J_a(x) J_a(y) f(x,y) ~+~ {c_A \over {\pi ^2}} {{{\rm dim}G} \over (x-y)^2} f(x,y)]
\eqno(4.10)
$$
is orthogonal to the ground state and obeys
$$
T~\Psi_2(J) ~=~ 2m~\Psi_2(J) \eqno(4.11)
$$
This state has invariance under $ H \rightarrow V H {\bar V}$ for constant $V, {\bar V}$.
Wavefunctions which have invariance for all (local) $V(z), 
{\bar V}({\bar z})$ can be obtained by taking appropriate limit; for example $f(x,y) = 
\bar \partial _x \bar \partial _y [\d (x-y) f(x)]$ would lead to $\bar \partial J_a (x) 
\bar \partial J_a (y)$ which is invariant under $H \rightarrow V(z) H \bar V (\bar z)$.
Given the result (4.8), it is clear that one can construct higher eigenfunctions
by proper orthogonalizations. Parity transforms of these states can also be obtained
by using ${\bar J_a}={(c_A/\pi) }(H^{-1}{\bar \partial}H)_a$. Each $J_a$ carries a spin of
$+1$, while ${\bar J}_a$ has a spin of $-1$. The normalization integral for these states
will involve the current correlators of the $G^{\bf C}/G$-WZW model and hence we expect
the states to be normalizable ( with suitable regularization, as is standard for states
in field theory). 

The requirement that we have eigenstates does not constrain the functions 
$f(x),~f(x,y)$, apart from normalization; we have an eigenstate for any 
$f$. Thus the states are infinitely degenerate. This degeneracy will
be lifted by the inclusion of the potential term; the effects of the potential should
be calculated using degenerate-state perturbation theory.
\vskip .1in
\noindent {\bf 5. Discussion}
\vskip .1in
We have used the parametrization of the gauge potential $A_z = -\partial
M~M^{-1}$, where $M$ is an $SL(N,{\bf C})$- matrix, to work out
the Hamiltonian formulation of an $SU(N)$ gauge theory in two spatial
dimensions. The matrix field $M$ is not gauge-invariant, but the theory
can be written in terms of $H= M^\dag M$ which is gauge-invariant.
The hermitian matrix field
$H$ can thus be taken as the basic field variable of the theory.
The simplicity of the parametrization we use is in the fact
that the relevant
Jacobian for the change of variables is easily evaluated in terms of the
WZW-action for the matrix field $H$. The volume of the configuration
space is then given by the partition function for the two-dimensional
WZW-theory for $H$. Within the standard requirements
of regularization ( such as limiting to a large but finite
number of modes), the volume is then finite; this is in contrast to
an Abelian theory where the volume so defined would be infinite.
This `finiteness' of the non-Abelian case is
very suggestive and is presumably related to
the mass gap, especially in view of intuitive
arguments outlined in the
introduction.

We have also obtained the Poincar\'e generators and done a direct check
of Poincar\'e invariance within our parametrization.
The kinetic energy term of the Hamiltonian is related to the Laplace
operator on the configuration space. If the `finiteness' of the 
volume of the configuration space is the
reason for the mass gap, one would expect to see it already at the level
of 
the spectrum of the kinetic term. The kinetic energy operator is especially
simple in our parametrization of fields (see eq.(3.18) or (4.6)).
The ground state wave function
for the kinetic operator is given by a constant and is
normalizable with the gauge-invariant measure.
We have obtained some
excited states which show a discrete spectrum, with a gap
$m= (e^2c_A/2\pi)$. The excited states are infinitely degenerate,
as should be expected, since they are eigenstates of only the
kinetic operator; the degeneracy will be lifted by the inclusion of
the potential term. 
Eventhough the Hamiltonian has a simple structure, clearly we do not
yet have a systematic calculational scheme. 
The proper inclusion of the effects of the potential
term and the full construction of the eigenstates are questions
which have to be addressed before meaningful calculations can
be attempted.
These issues are under investigation. 
\vskip .2in
We thank R. Jackiw, B. Sakita and E. Witten for useful discussions.
Discussions with
G. Alexanian, A. Kavalov, Chanju Kim and D. Minic are also gratefully acknowledged. 
Special thanks are due to S. Samuel for pointing out a regularization ambiguity in
a previous version of this paper.
This work was supported in part by the Department of Energy, grant numbers
DE-FG02-90ER40542 and DE-FG02-91ER40651-Task B and the National Science Foundation,
grant number PHY-9322591.
\vskip .1in
\noindent{\bf Appendix}
\vskip .1in
In parametrizing the fields as in eq.(2.4) and using the gauge-invariant variable
$H$, one of the potential difficulties that one may worry about is the
question of coordinate singularities.
The physical configuration space ${\cal C}$ 
is the space of gauge potentials ${\cal A}$ modulo ${\cal G}_*$, the latter
being the set of gauge transformations which go to the identity at spatial infinity.
${\cal A}$, considered as a ${\cal G}_*$-bundle over ${\cal C}$ is nontrivial.
Thus one cannot choose global sections; this is the well known Gribov problem [13].
The space ${\cal C}$
is topologically and geometrically
nontrivial.

The simplest way to see the nontriviality of ${\cal A}$ as a ${\cal G}_*$-bundle is to
consider the homotopy groups. Since ${\cal A}$ is homotopically trivial, if we show that
${\cal C}$ has nontrivial homotopy groups, it is clear that ${\cal A}$ cannot be written 
as a product ${\cal C}\times {\cal G}_*$. Since $\Pi_1 ({\cal G_*}) \approx
\Pi_3 (G) ={\bf Z}$, it is easily checked that $\Pi_2 ({\cal C})={\bf Z}$. This implies 
that there are
noncontractible closed two-surfaces in ${\cal C}$. This is the simplest obstruction to 
the triviality of the bundle [13].
The existence of topologically
nontrivial structures in the configuration space implies that any gauge-invariant
parametrization of the fields or choice of coordinates on ${\cal C}$ will
necessarily have coordinate singularities [14]. 
We may hope to gain 
some understanding of the importance of the coordinate singularities
by studying configurations which form a noncontractible two-surface.

It is not too difficult to construct a set of configurations which form a noncontractible
two-surface since they are related to the instanton of the four-dimensional
gauge theory. This can be seen as follows.
In addition to $\Pi_2({\cal C})$ being nontrivial,
the second cohomology group of ${\cal C}$ is nontrivial as well. In other
words, there is a closed but not exact two-form on ${\cal C}$. In terms of the
potentials, a representative of this can be written as
$$
\Omega ={1\over 4\pi} \int \Tr (\delta A ~ \delta A) \eqno(A.1)
$$
The integral of $\Omega$ over the closed noncontractible two-surface in ${\cal C}$ gives
a winding number $\nu$ by
$\int \Omega ~=2\pi \nu $. The two-surface in ${\cal C}$ along with the two-dimensional
spatial manifold gives
a four-dimensional space and $\nu$ is the instanton number on this space [15]. 
Specifically,
$$
\nu ~= {1\over 8\pi ^2} \int \Tr ({\tilde F} ~{\tilde F}) \eqno(A.2)
$$
where ${\tilde F}= (d+\delta ) {\tilde A} +{\tilde A}{\tilde A}$. 
Here ${\tilde A}$ is the
four-dimensional gauge potential; it is constructed from the two-dimensional potential $A$
as ${\tilde A}= A +c_w dw +c_{\bar w}
d{\bar w}$ and $\delta = dw \partial_w + d{\bar w}\partial_{\bar w}$. 
$c_w,~c_{\bar w}$ are given in terms of $M$ and $M^\dag$ by
$$
c_w =- \partial_w M ~M^{-1},~~~~~~~~~~c_{\bar w}= (M^\dag)^{-1}\partial_{\bar w}M^\dag
\eqno(A.3)
$$
In terms of these variables
$$
\Omega ={1\over 2\pi} \int \Tr \left[ \partial (H^{-1}\bdel H)\delta (H^{-1} {\bar \delta}
H)~+\partial (H^{-1}{\bar \delta} H) \delta (H^{-1}\bdel H)\right]\eqno(A.4)
$$
We can exploit the connection outlined above between the two-form
$\Omega$ on ${\cal C}$ and the instanton number to construct an example
of the noncontractible
two-surface of configurations.
The standard instanton in ${\bf R}^4$ [16] can be rewritten using complex coordinates
and interpreting one pair of complex coordinates as internal coordinates, viz.,
as parametrizing the two-surface in ${\cal C}$, we can get a set of  
configurations of interest.
Explicitly we have
$$
H= \exp (2f J^3) ~=\cosh 2 f ~+J^3 \sinh 2f \eqno(A.5)
$$
Here $J^3 =\sigma \cdot n $; $\sigma^a,~a=1,2,3$, are the Pauli matrices and the unit
vector $n^a$ is
given by
$$
n^a={1\over ({\bar z}z+{\bar w}w)}
\left ( {\bar z} w+{\bar w}z, ~i({\bar w}z -{\bar z}w),~ {\bar z}z-{\bar w}w \right)
\eqno(A.6)
$$
Also
$$
f ~= \12 \log \left( {{\bar z}z+{\bar w}w +\mu^2 \over {{\bar z}z+{\bar w}w}}\right)
\eqno(A.7)
$$
$\mu$ is a scale parameter and $(w, {\bar w})$ parametrize the two-surface
in ${\cal C}$. It is easily verified that $\nu =\int {\Omega /2\pi }$ is equal to
3 for this set of configurations and hence eq.(A.5) gives a noncontractible two-surface
in ${\cal C}$. As ${\bar z}z \rightarrow \infty,~H\rightarrow 1$. For almost all 
$w, {\bar w}$, $H$ is nonsingular; however, the configuration at $w=0$ has a singularity
at the spatial point $z=0$. One can shift the position of this singularity
by transformations of the type $H\rightarrow V H{\bar V}$, where $V$ is holomorphic in
$z$. Nonsingular configurations are given by nonsingular formulae for 
$H$ in different coordinate patches with transition relations given by
transformations of this type. Since the singularity
in our example is at a point, viz., at $w=0$, even if we simply use the formulae
(A.5-7) with the coordinate singularity, the effect on the quantum wave functions
is minimal; one can see this explictly by constructing wave functions $\psi (w,{\bar w})$
for the reduced set of configurations (A.5). We can also consider the effect on the
vacuum wave function $\Phi_0$. The WZW-action is invariant under transformations of the 
type $H\rightarrow V H{\bar V}$ and we therefore
do not expect any pathology for the wave function.
Explicitly, for the set of configurations (A.5), the WZW-action is given by
$$
{\cal S} (H) ~= {{ 5 \mu^2 +4w{\bar w}}\over {w{\bar w}+\mu^2}} ~-
{{ 3 \mu^2 +4 w{\bar w}}\over{\mu^2}} \log \left[ {{ \mu^2 +w{\bar w}} \over
{w {\bar w}}}\right] \eqno(A.8)
$$
When $w\rightarrow 0$, $\Phi_0 = \exp (c_A \S)$ vanishes as $(w{\bar w})^{3c_A}$. 
The coordinate singularity does not
lead to difficulties, at least for this case, and a
nonsingular description is not essential.
\vskip .2in
\noindent
{\bf References}
\vskip .1in
\item
{1.} A.M. Polyakov, {\it Nucl.Phys.} {\bf B120 } (1977) 429.
\vskip .1in
\item
{2.} G.'t Hooft, {\it Nucl.Phys.} {\bf B138} (1978) 1;
R. Jackiw and S. Templeton, {\it Phys.Rev.} {\bf D23} (1981) 2291;
J. Schonfeld, {\it Nucl.Phys.} {\bf B185} (1981) 157; S. Deser, R. Jackiw and
S. Templeton, {\it Phys.Rev.Lett.} {\bf 48} (1982) 975; {\it Ann.Phys.}
{\bf 140} (1982) 372.
\vskip .1in
\item {3.} J. Goldstone and R. Jackiw, {\it Phys.Lett.} {\bf 74B} (1978) 81; M.B. Halpern, {\it Phys. Rev.} {\bf D16} (1977) 1798; {\it ibid.} {\bf D16} (1977) 3515; {\it ibid.} {\bf D19} (1979) 517; I. Bars and F. Green, {\it Nucl. Phys.} {\bf B148} (1979) 445.
\item
{4.} D.Z. Freedman, {\it et al}, MIT preprint hep-th 9309045; 
D.Z. Freedman and R. Khuri, {\it Phys.Lett.}
{\bf 192A} (1994) 153; M. Bauer and D.Z. Freedman, MIT preprint hep-th 9505144;
F.A. Lunev, {\it Phys.Lett.} {\bf 295B} (1992) 99; {\it Theor.Math.Phys.}
{\bf 94} (1993) 66; {\it Mod.Phys.Lett} {\bf A9} (1994) 2281; M. Asorey, {\it et al},
preprint hep-th 9502024; I. Kogan and A. Kovner, {\it Phys.Rev.} {\bf D51} (1995)
1948; S.R. Das and S. Wadia, preprint hep-th 9503184; O. Ganor and J. Sonnenschein,
preprint hep-th 9507036.
\vskip .1in
\item{5.} R.P. Feynman, {\it Nucl.Phys.} {\bf B188} (1981) 479; I.M. Singer,
{\it Phys.Scripta} {\bf T24} (1981) 817.
\vskip .1in
\item
{6.} J. Schwinger, {\it Phys.Rev.} {\bf 122} (1962) 324; {\it ibid.} {\bf 130} (1963)
406; J.L. Gervais and B. Sakita, {\it Phys.Rev.} {\bf D18} (1978) 453; B. Sakita,
{\it Phys.Rev.} {\bf D21} (1980) 1067; in {\it Field Theory in Elementary
Particles}, B. Kursunoglu and A. Perlmutter (eds.) (Plenum, 1983); 
N. Christ and T.D. Lee, {\it Phys.Rev.} {\bf D22} (1980) 939.
\vskip .1in
\item
{7.} E. Witten, {\it Commun.Math.Phys.} {\bf 92} (1984) 455;
S.P. Novikov, {\it Usp.Mat.Nauk} {\bf 37} (1982) 3.
\vskip .1in
\item
{8.} D. Karabali, Q-H. Park,
H.J. Schnitzer and Z. Yang, {\it Phys.Lett.} {\bf 216B} (1989) 307;
D. Karabali and H.J. Schnitzer, {\it Nucl.Phys.} {\bf B329} (1990) 649.
\vskip .1in
\item
{9.} K. Gawedzki and A. Kupiainen, {\it Phys.Lett.} {\bf 215B} (1988) 119;
{\it Nucl.Phys.} {\bf B320} (1989) 649.
\vskip .1in
\item
{10.} M. Bos and V.P. Nair, {\it Int.J.Mod.Phys.} {\bf A5} (1990) 959.
\vskip .1in
\item
{11.} A.M. Polyakov and P.B. Wiegmann, {\it Phys.Lett.} {\bf 141B} (1984) 223.
\vskip .1in
\item
{12.} R. Efraty and V.P. Nair, {\it Phys.Rev.} {\bf D47} (1993) 5601;
 V.P. Nair, {\it Phys.Rev.} {\bf D48} (1993) 3432.
\vskip .1in
\item
{13.} V. Gribov, {\it Nucl.Phys.} {\bf B139} (1978) 1; I.M. Singer, {\it Commun.Math.Phys.}
{\bf 60} (1978) 7; T. Killingback and E.J. Rees, {\it Class.Quant.Grav.} {\bf 4}
(1987) 357.
\vskip .1in
\item 
{14.} W. Nahm, in {\it Proceedings of the IV Warsaw Symposium on Elementary
Particle Physics}, Z. Adjuk (ed.) (Warsaw, 1981). 
\vskip .1in
\item
{15.} V.P. Nair and J. Schiff, {\it Nucl.Phys.} {\bf B371} (1992) 329.
\vskip .1in
\item
{16.} see for example, T. Eguchi, P. Gilkey and A. Hanson, {\it Phys.Rep.} {\bf 66} (1980)
213.

\end